\documentstyle[12pt]{article}
\title{THE QCD STRING WITH QUARKS. II. LIGHT CONE HAMILTONIAN}
 \author{ A.Yu.Dubin, A.B.Kaidalov and Yu.A.Simonov\\
Institute of Theoretical and Experimental Physics\\ 117259, Moscow,
B.Cheremushkinskaya 25, Russia}
\date{}
\newcommand{\be}{\begin{equation}}
\newcommand{\ee}{\end{equation}}
\begin{document}
\maketitle
\begin{abstract}

The light-cone Hamiltonian is derived from the general gauge -- and Lorentz
-- invariant expression for the $q\bar{q}$ Green's function, containing
confinement via the area law for the Wilson loop.
The resulting Hamiltonian contains in a nonadditive way contributions from
quark and string degrees of freedom. Different limiting dynamical regimes of
the system of quarks connected by the string are found. In the limit of no
transverse motion one recovers the t'Hooft's 1+1 QCD Hamiltonian. The
correspondence with the rest frame  Hamiltonian spectrum is established.
The important notion of the light-cone string, contributing significantly to
the energy--momentum is discussed.
  \end{abstract}
   \newpage
\section{Introduction}

\large
We have studied recently ([1], hereafter denoted as I) the quark-antiquark
system in the confining vacuum. Starting from the QCD Lagrangian and
assuming the minimal area law for the Wilson loop we have derived the
dynamics of the system, where in addition to quark and antiquark also the
QCD string appears at large  distances, $R>T_g$, where $T_g$ is the
correlation length of the gluoic  vacuum.

We have found in I the Hamiltonian of the system in the c.m. frame, which
contains degrees of freedom of quarks and those of the string. The latter
enters via an auxiliary function $\nu$, introduced in I, which physically
measures the energy density of the string.

The Hamiltonian describes two different  limiting regimes (for light
quarks); for low orbital momentum $ n_r \gg L$ one obtains the relativistic
linear potential
Hamiltonian [2] describing radial excitations, while for high
$L\gg n_r$ one gets the regime of a rotating string.

It is remarkable that when correction terms are taken into account both
regimes yield [3] almost the same standard slope
$\alpha'= (2\pi\sigma)^{-1}$ of Regge trajectories.
This  outcome also agrees with the result of numerical quantization in [4].
Our derivation in I essentially exploited the c.m. frame, and we have
neglected in the path integral the backtracking paths of the quark and
antiquark, arguing that those are suppressed on dynamical grounds.

The light-cone frame is advantageous from several points of view. First, the
light-cone dynamics is believed to be simpler, especially for perturbative
contributions [5]. Second, the backtracking paths do not contribute on the
light cone for kinematical reasons. Last, but not the least, the light-cone
formalism based on light-cone Hamiltonian, allows one to make contact with
the parton picture in QCD [6] and to calculate formfactors and structure
functions in terms of the eigenfunctions of the light-cone Hamiltonian.

Recently a promising formalism was proposed and shown to be effective at
least for treatment of perturbative interaction -- the light cone
quantization [7]. Our light-cone Hamiltonian which will be obtained  below in
section 2
 contains also the nonperturbative interaction between quarks -- the
string, and one can test on it methods of [7] or modification of those.

It is a purpose of the present paper to apply the method of I in the
light-cone  frame and to calculate the light-cone Hamiltonian for the system
of quark and antiquark in the nonperturbative (confining) vacuum. As in I,
we exploit the formalism of the Feynman-Schwinger representation for the
$q\bar{q}$ Green's function [8,9] together with the Hamiltonian formalism of
[10,9]. We neglect for simplicity quark pair creation and perturbative gluon
exchanges, concentrating only on the nonperturbative interaction, producing
(initially in Euclidean space) the minimal area surface, bounded
by paths of quark and antiquark.  The Hamiltonian we are looking
for, depends on the chosen evolution parameter $\tau$ and is defined after
continuation into Minkowski space on the 3-hypersurface (light cone plane)
in $4d$.

The quantum Hamiltonian is in general both first and second quantized, i.e.
is an operator in terms of field creation-annihilation operators and also an
operator in space of  coordinates of particles and antiparticles.

Since as in I  we use the minimal area law asymptotics (which corresponds to
neglect of string  excitations ) and  neglect quark pair creation
together with  perturbative gluon exchanges there is a possibility
to obtain only first-quantized Hamiltonian. The price paid in I
was that backtracking paths had to be neglected too, since they
correspond (after approximations beeing made) to pair creation processes
from the vacuum.  We find below the light-cone Hamiltonian , which is again
only first-quantized, and additional neglect of no backtracking is not
necessary since it is contained automatically in the light-cone frame.

Thus the light-cone dynamics also reduces to the relativistic quantum
mechanics of quarks connected by the string, as in I, but without additional
assumptions.

As well as in I we find two dynamical regimes of the "minimal" QCD string
(corresponding to the minimal area law asymptotics of the averaged Wilson
loop): for large orbital excitations $L\approx |L_z|\gg n_r $ the mass
squared $M^2$ and the total momentum $P_+$ are determined by the rotating in
transverse plane  string contribution; for large radial quantum numbers
$n_r \gg L$ one gets the relativistic potential Hamiltonian so that the
string contribution is "inert" in the leading order.  We find the agreement
between
the calculations of the spectrum in the limit $L\to \infty $ or $n_r\to
\infty$ in the rest and light--cone frames, postponing the detailed
comparison of the Regge trajectories in I and II till the next publication.

We postpone the detailed discussion of relation of our Hamiltoian to the
standard relativistic quantum mechanics formalism [11],[12] based on the
Dirac paper [11] till the next publication. We note only that usual
assumption, that the interaction doesn't contribute to the total momentum
$P_+$ and to the orbital momentum is not valid for our system because the
string carries the part of $P_+$.

The paper is organized as follows. In section 2 the light-cone Hamiltonian
is derived from our general expressions in I. In Section 3 we study the
spectrum of the obtained Hamiltonian and elaborate in detail the limiting
 cases such as the t'Hooft's 1+1 QCD at $N_C\rightarrow\infty$, the
 heavy quarkonia case and three  limiting relativistic regimes.
Section 4 is devoted to summary and conclusions.
In Appendix we establish a connection between our approach and t'Hooft 1+1
QCD equation.

\setcounter{equation}{0}
\renewcommand{\theequation}{2.\arabic{equation}}

\section{The light-cone Hamiltonian}

Given a $q\bar{q}$ Green's function in the coordinate space
$G(x\bar{x};y\bar{y})$, where $x\bar{x}(y\bar{y})$ are final (initial)
4-coordinates of quark and antiquark, one can define the Hamiltonian $H$
through the equation (in the Euclidean space-time)
\be
\frac{\partial G}{\partial T} = - HG
\ee
where $T$ is an evolution parameter corresponding to some choice of a
$3d$ hypersurface $\Sigma$. In the particular case of the c.m. Hamiltonian
in I the role of $T$ is played by the center-of-mass Euclidean time
coordinate $T=\frac{x_4+\bar{x}_4}{2}$ and the hypersurface $\sum$ is a
hyperplane $x_4=\bar{x}_4=const.$

In general, $H$ is a  i) second -- quantized and a ii) first -- quantized
operator, which means that  i) $H$ has matrix elements connecting different
number of particles
 and ii) $H$ is an operator acting on the
coordinates of each particle.

In our case, we neglect all perturbative gluon exchanges (small $\alpha_s$
at all distances [13]) and quark-pair creation (small $1/N_C$) and  keep
only nonperturbative confining interaction leading in Euclidean
space to the area law of the Wilson loop. With all that, the
first-quantized Hamiltonian can be defined only at large distances $R\gg
T_g$ [14] where $T_g$ is the vacuum correlation length [15].

Thus the problem  becomes that of relativistic quantum mechanics and in I we
have found the relativistic  Hamiltonian in the c.m. frame.

In this paper we aim at finding Hamiltonian in the light-cone frame, which
can be done directly via (2.1) and the Feynman-Schwinger representation of
the $q\bar{q}$ Green's function [8,9]. This type of analysis in a more
general case, when also perturbative gluons are taken into account, will be
a subject of another publication, and here instead we shall follow the
formalism of I, adjusting it to our specific frame.

With the notations  for the vectors $a_{\mu},b_{\mu}$
\be
ab= a_{\mu}b_{\mu} = a_ib_i-a_0b_0=a_{\bot}b_{\bot}+a_+b_-+a_-b_+,
\ee
$$a_{\pm}=\frac{a_3\pm a_0}{\sqrt{2}},$$
one can define the hypersurface $\sum$ through the $q\bar{q}$ coordinates
$z_{\mu}.\bar{z}_{\mu}$ as
\be
z_+(\tau)=\bar{z}_+(\bar{\tau})
\ee
and the kinetic part $K+\bar{K}$ of the action $A$, (I,eqs(25-26)).  \be
A=K+\bar{K}+\sigma S_{min},
\ee
has the form (after the continuation of eq.(22) into the Eucledian space)
\be
K+\bar{K} = \frac{1}{4} \int^s_0 \dot{z}^2_{\mu}(\tau)d\tau+\frac{1}{4}
 \int^{\bar{s}}_0 \dot{\bar{z}}^2_{\mu}(\bar{\tau})d\bar{\tau}+
  \int^{s}_0m_1^2 d\tau+ \int^{\bar{s}}_0m_2^2 d\tau
  \ee
  $$=
  \int^T_0dz_+[\frac{\mu_1}{2}(\dot{z}^2_{\bot}+2\dot{z}_-)+\frac{\mu_2}{2}
  (\dot{\bar{z}}^2_{\bot}+2\dot{\bar{z}}_-)
  +\frac{m_1^2}{2\mu_1}+\frac{m^2_2}{2\mu_2}]$$
  where instead of proper time $\tau$ we have used $z_+$ and
  introduced new variables $\mu_1, \mu_2$ via
   \be
  2\mu_1(z_+)=\frac{d z_+}{d \tau}~;~~
  2\mu_2(z_+)=\frac{d \bar{z}_+}{d \tau}~; ~~z_+=\bar{z}_+
  \ee
  The condition of no backtracking $\dot{\bar{z}}_+, \dot{z}_+>0$ discussed
  above implies $\mu_1>0, \mu_2>0$.
  Dots in (2.5)  imply derivatives in $z_+$,
  while \be T=\frac{x_++\bar{x}_+}{2}=x_+ \ee

  For the minimal area  surface $S_{min}$ we use the approximation that it
  is the worldsheet of the straight line connecting $z_{\mu}(z_+)$ and
  $\bar{z}_{\mu}(z_+)$ with the same value of the evolution parameter $z_+$,
  i.e.
    \be S_{min}=  \int^T_0 dz_+\int^1_0 d\beta [\dot{w}^2
  w'^2-(\dot{w}w')^2]^{1/2} \ee where \be w_{\mu}(z_+; \beta) = z_{\mu}(z_+)
  \beta + \bar{z}_{\mu}(z_+) (1-\beta)
   \ee
   and dot and prime denote
  derivatives  in $z_+$ and $\beta$ respectively.

  This approximation corresponds o the instantoneous formation of the string
  (flux tube) in accordance with the positions of quarks as in [10,1]. We
  now introduce "center-of-masses" and relative coordinates,(see[16] for a
  detailed discussion)
   \be
  \dot{R}_{\mu}=x(\tau)\dot{z}_{\mu}+(1-x(\tau))\dot{\bar{z}}_{\mu}~,~~
  \dot{r}_{\mu}=\dot{z}_{\mu}-\dot{\bar{z}}_{\mu}~
  \ee
  where function $x(\tau)$ will be determined below and also
  auxiliary functions $\eta(\beta, x_+),\nu(\beta,x_+)$ as in I will enter
  our action  to get rid of the square root in (2.8).

  Similarly to I we obtain
  \be
G(x\bar{x};y\bar{y})= \int D{\mu}_1(z_+)D\mu_2(z_+)D\nu\cdot D\eta
DR_{\mu}Dr_{\mu}e^{-A}
\ee
where
\begin{eqnarray}
A=\frac{1}{2}\int^T_0 dz_+\{a_{1}\dot{R}^2_{\bot}+2a_1\dot{R}_- -
2c_1\dot{R}_{\bot}r_{\bot}+2a_2\dot{R}_{\bot}\dot{r}_{\bot}+
\\
\nonumber
+2a_2\dot{r}_- -2c_2\dot{r}_{\bot}r_{\bot}+a_3\dot{r}^2_{\bot}-
2c_1r_-+a_4r^2_{\bot}+\frac{m_1^2}{\mu_1}+\frac{m^2_2}{\mu_2}\}
\end{eqnarray}
and we have used notations
\be
a_1=\mu_1+\mu_2+\int^1_0\nu(z_+,\beta)d\beta
\ee
\be
a_2=\mu_1(1-x)-x\mu_2+\int^1_0d\beta \nu(1-x\beta-x(1-\beta))
\ee
\be
a_3=\mu_1(1-x)^2+\mu_2x^2+\int^1_0 \nu(z_+,\beta)(\beta-x)^2 d\beta
\ee
\be
a_4=\int^1_0 \frac{\sigma^2}{\nu}d\beta+\int^1_0\eta^2\nu d\beta
\ee
\be
c_1=\int^1_0 \eta\nu d\beta
\ee
\be
c_2=\int^1_0 \eta\nu (\beta-x) d\beta
\ee

One can determine now  $x(\tau)$ from the requirement that
 $\dot{R}$ is to be decoupled from $\dot{r}$ to provide the
diagonalization [16] of the quadratic velocity part of the action  (2.12).
Formally it results in the condition $a_2=0$ which gives
\be
x=\frac{\mu_1+\int \nu\beta
d\beta}{a_1};~~ 1-x=\frac{\mu_2+\int \nu(1-\beta) d\beta}{a_1};~~ \ee
As we shall see from Section 3 function $x(\tau)$ plays the role of the
Feinman light cone variable.

Integration over $\eta$ in (2.11) with $A$ given by (2.12) can
be done in the same way as in I and one gets a new effective action $A'$
(here and in what follows we take into account only the exponent and not the
preexponential factor, since we are interested in terms in the exponent
linearly growing with $T$). We have \normalsize
\begin{eqnarray}
A'=\frac{1}{2}\int dz_+\{\frac{m^2_1}{\mu_1}+\frac{m^2_2}{\mu_2}+
a_1(\dot{R}^2_{\bot}+2\dot{R}_-)+a_3\dot{r}^2_{\bot}+
\int\frac{\sigma^2}{\nu}d\beta \cdot r^2_{\bot}\\
\nonumber
- \frac{(r_-+\dot{R}_{\bot}r_{\bot}+(<\beta>-x)\dot{r}_{\bot}r_{\bot}
)^2}{r^2_{\bot}(\int \nu d\beta)^{-1}}-
\frac{(\dot{r}_{\bot}r_{\bot})^2\int\nu(\beta-<\beta>)^2d\beta}{r^2_{\bot}}\},
\end{eqnarray}
\large
where we define $<\beta>=\int\nu\beta d\beta/\int\nu d\beta$.

We now proceed as in I to integrate over $D\dot{R}_{\mu}$ instead of $DR$
using relations
\normalsize
\be
\int DR e^{-A'}= \int D\dot{R}\int d^3\lambda e^{-A'}\cdot e^{i\lambda_{\bot}
\int^T_0\frac{d{R}_{\bot}}{d\tau}dz_+
+i\lambda_+\int
\frac{d{R}_-}{d\tau}dz_+}\cdot
\ee
$$
\cdot e^{-i\lambda_{\bot}(R_{\bot}(T)-R_{\bot}(0))}
 e^{-i\lambda_{+}(R_-(\tau)-R_{-}(0))}
$$

\large
Fixing the frame of reference we  finally have  $\lambda_{\mu}\to
P_{\mu}$, where $P_{\mu} $ is total c.m. momentum so that one obtains for
$\vec{P}_{\bot}=0$ the new effective action
 \begin{eqnarray}
A^{\prime\prime}= \frac{1}{2} \int^T_0 dz_+ \{ \frac{m^2_1}{\mu_1}+
\frac{m^2_2}{\mu_2}+
a_3\dot{r}^2_{\bot}+\int
\frac{\sigma^2}{\nu} d\beta \cdot r^2_{\bot} -\\
\nonumber
\nu_2
\frac{(\dot{r}_{\bot}r_{\bot})^2}{r^2_{\bot}}-
\frac{\nu_0 a_1 [r_-
+(<\beta>-x)\dot{r}_{\bot}r_{\bot}]^2}{r^2_{\bot}(\mu_1+\mu_2)} \}
\end{eqnarray}
where $\nu_k=\int \nu(\beta, z_+)(\beta-<\beta>)^kd \beta$. Integration over
$D\dot{R}_-$ yields $\delta(a_1-P_+)$ with an important constraint
\be
a_1=P_+
\ee
while integration over $DR_+$ is trivial since $A'$ does not depend on
$\dot{R}_+$.

It is convenient to represent the constant (2.23) (existing for any point
$z_+$,
so that we have $\pi_i\delta(a_1(z_+^i)-P_+))$ introducing the following
representation of $\delta$ -- function \be \delta(a_1(z_+)-P_+)=
\frac{1}{2\pi} \int D\alpha(z_+) e^{i\alpha(z_+)(a_1(z_+)-P_+)} \ee

Furthermore we go over into the Minkowski space, which means that
$$\mu_i\to - i\mu^M_i~,~~\nu \to -i\nu^M$$
\be
a_i\to -ia^M_i, ~~A\to -i A^M
  \ee
 For the Minkowski action we obtain  (omitting from now on the superscript
 $M$ everywhere)
 \normalsize
 \begin{eqnarray}
 A^M&=& \frac{1}{2} \int dz_+\{
 -\frac{m^2_1}{\mu_1}-\frac{m^2_2}{\mu_2}+a_3\dot{r}^2_{\bot}- \int
 \frac{\sigma^2 d\beta }{\nu}r^2_{\bot}-\\
 \nonumber
& -&\nu_2\frac{(\dot{r}_{\bot}r_{\bot})^2}{r^2_{\bot}}-
\frac{\nu_0a_1}{(\mu_1+\mu_2)r^2_{\bot}}[r_-+(<\beta>-x)\dot{r}_{\bot}
r_{\bot}]^2 +2\alpha(a_1-P_+)\}
 \end{eqnarray}

\large
 Let us now define the Hamiltonian, corresponding to the action $A^M$
 \be
 A^M= \int dz_+ L^M~, ~~H=p_{\bot}\dot{r}_{\bot}-L^M
 \ee
 with
 $p_{\bot}=\frac{\partial L^M}{\partial \dot{q}_{\bot}}$.
 Simple calculations lead to the following expression
 \begin{eqnarray}
 H=\frac{1}{2} \{ \frac{m^2_1}{\mu_1}+\frac{m^2_2}{\mu_2}+
 \frac{p^2_{\bot}-\frac{(p_{\bot}r_{\bot})^2}{r^2_{\bot}}}{a_3}+
 \frac{(p_{\bot}r_{\bot}+\gamma r_-)^2}{\tilde{\mu} r^2_{\bot}}\\
 \nonumber
 +\int \frac{\sigma^2}{\nu} d\beta r^2_{\bot}+ \frac{\nu_0 a_1}{\mu_1+\mu_2}
 \frac{r^2_-}{r^2_{\bot}}  \}
 \end{eqnarray}
 where we have used notations
 $$~~ \gamma
 =\nu_0(<\beta>-\frac{\mu_1}{\mu_1+\mu_2})= a_1(x-\frac{\mu_1}{\mu_+}),
{}~~\tilde{\mu}=\frac{\mu_1\mu_2}{\mu_1+\mu_2}
$$
 and the condition (2.23) is implied.

We note that additionally to canonically conjugated pairs $\{ x,(P_+r_-)\}$
 (see Section 3 for a short discussion), $\{\vec{p}_{\bot},
 \vec{r}_{\bot}\}$ Hamiltonian (2.28) contains the auxiliary field
 $\nu(\tau,\beta)$ playing the role of the string energy density. Variables
 $\mu_1, \mu_2, a_3$ are to be expressed with the help of eqs. (2.19),
 (2.23) in terms of $x$ and $\nu$.

  The integration over $\nu(\tau, \beta)$ effectively amounts [1] to the
 substitution into eq. (2.28) its extremal value
 \be
  \frac{\delta
 H}{\delta\nu(\tau,\beta)}\mid_{\nu=\nu^{ext}}=0
 \ee
 Only after this
  substitution one is to construct the operator Hamiltonian acting on the
 wave functions  (for more detailed discussion see forthcoming paper).

    The Hamiltonian (2.28)
 is the central result of our paper.  In the next section we shall study its
 properties.

\setcounter{equation}{0}
\renewcommand{\theequation}{3.\arabic{equation}}

\section{Properties of the light-cone Hamiltonian}

We consider in this section  the case of heavy quarks and then different
limiting  relativistic regimes of the light--cone Hamiltonian.

\subsection{Heavy quark limit of light--cone Hamiltonian}

Assuming that quark masses are large,
\be
m_1\gg\sqrt{\sigma}, ~~
m_2\gg\sqrt{\sigma}, ~~
\ee
one can prove that $\nu$ does not depend on $\beta$ in the leading order.
Let us introduce the variable $y=\frac{\nu}{P_+}$ corresponding to the part
of the total momentum carried by the string. We shall demonstrate that the
effective values of $y$ satisfy the condition
\be
 y\ll 1
 \ee
 First one is to
expand kinetic part of (2.28) around the extremal value of $x$
\be
 x_{ext} =\frac{m_1}{m_1+m_2} \ee
  with the result
  \be
(m_1+m_2)^2+2(m_1+m_2)\frac{1}{2\tilde{m}}(\vec{p}^2_{\bot}+p^2_z)+(m_1+m_2)^2y
\ee
where $\tilde{m}=m_1m_2/(m_1+m_2)$ and we defined
\be
p_z\equiv (m_1+m_2)(x-m_1/(m_1+m_2))
\ee
We stress here that kinetic part of eq. (2.28) actually contains not only
the analog of the rest frame kinetic term $\frac{\vec{p}^2}{2\tilde{m}}$,
but also the part of the potential $(m_1+m_2)y$. Therefore the Hamiltonian
(2.28) can not be represented in relativistic case as a sum of pure kinetic
and pure potential term as it is assumed in some approaches [12].

Substituting (3.4) into (2.28) one arrives at the Hamiltonian in the form \,
where auxiliary function $y$ participates
\begin{eqnarray}
H=\frac{1}{2P_+}\{(m_1+m_2)^2+2(m_1+m_2)[\frac{1}{2\tilde{m}}(p^2_{\bot}
+p^2_z)+
\\
\nonumber
+(m_1+m_2)\frac{1}{2}(\frac{y}{r_{\bot}^2})(r^2_{\bot}+
(\frac{(P_+r_-)}{m_1+m_2})^2+
\frac{\sigma^2}{2(m_1+m_2)}(\frac{r^2_{\bot}}{y})]\}
\end{eqnarray}
The integration over $y$ in the path integral with the Hamiltonian (3.6)
 amounts [1] to the insertion of the extremal value
of $y$
\be
y_{exp}=\frac{\sigma
r^2_{\bot}}{m_1+m_2}\cdot(r^2_{\bot}+(\frac{(P_+r_-)}{m_1+m_2})^2)^{-1/2}
\ee
and the condition (3.2) is indeed satisfied.

Finally introducing $r_z$, canonically conjugated to $p_z$ (3.16) via
relation
\be
r_z\equiv \frac{(P_+r_-)}{m_1+m_2}
\ee
one obtains the  Hamiltonian of heavy quarkonia  in the light--cone
system:
\be
H_{nonrel}=\frac{1}{2P_+}
\{(m_1+m_2)^2+2(m_1+m_2)(\frac{1}{2\tilde{m}}\vec{p}^2+\sigma |\vec{r}|)\}
\ee
or since $H_{nonrel}=\frac{1}{2P_+}M^2_{nr}$, one has in the leading order
\be
M_{nr}\cong m_1+m_2+\frac{1}{2\tilde{m}}\vec{p}^2+\sigma|\vec{r}|,
\ee
in agreement with the usual nonrelativistic result in the c.m. system.
We note also that variable $(x-\frac{m_1}{m_1+m_2})$ is canonically
conjugated to $(P_+r_-)$, which follows from eqs. (3.5), (3.8).

We emphasize that after elimination of the auxiliary function the 0(3)
rotation invariance of the Hamiltonian is recovered, which ensures that our
straight--line
anzatz on light--cone is compatible with the conservation of the angular
momentum.

\subsection{Relativistic dynamical regimes on light-cone}

To consider relativistic regimes of the "minimal" QCD string with quarks
in terms of light--cone variables, it is profitable to make correspondence
with the dynamics of the system in the rest frame where as we proved in I
there exist two limiting relativistic regimes. For the case of large orbital
excitations (case 1 in what follows) $L\gg n_r$ ($n_r$ is the radial quantum
number) the system behaves as the rotating string, which carries the main
part of the orbital momentum and the energy. In the opposite case
$n_r\gg L$ (case 2) the string is nearly pure inert (constituting in the
leading order the linear potential) and almost doesn't contribute into the
kinetic part of the Hamiltonian. We have found in I that the spectrum is
very close to the following asymptotic form
 \be M^2_n= 2\pi\sigma (2n_r+L+const) \ee
When one goes over to the light--cone, there appears as we will
demonstrate three  different dynamical regimes, which can be connected to
the two aforementioned of the  rest frame ones by an infinite  boost along
 $z$--axis.  To establish this relation  we are to recover first  the
counterpart of the rest frame $r_z$ on the light cone. From the expression
(3.8) for the heavy  quarkonia limit of the Hamiltonian (2.28) one can
anticipate, that it is the combination
 \be
 (P_+r_-)/M
 \ee
 where $M$ is the total meson mass which corresponds to the rest frame
 $r_z$ component. We note here that expression (3.12) obviously takes into
account well known Lorentz squeezing of longitudinal size for moving objects
and determines the proper longitudinal scale of the light cone dynamics.

Due to the fact, that $z$-axis is distinguished from transverse
ones the limiting dynamical regimes on the light cone are realized either
for stretched configuration \be \frac{|P_+r_-|}{M}\gg |r_{\bot}| \ee or for
squeezed into the perpendicular plane one \be \frac{|P_+r_-|}{M}\ll
|r_{\bot}| \ee

To form the corresponding wave packets in the rest frame one is to use for a
given $L\gg 1$ the set of adjoint Legendre polynomials $P_L^{L_z}(cos
\theta)$, $|L_z|\leq L$.
 If $\Delta \theta$ is angular smearing of the configuration
with respect to  $z$-axis, then the stretched wave packet,
$|r_z|\gg|r_{\bot}|$ is constructed from a set of $n\sim 1/\Delta\theta$
polynomials $P_L^{L_z}$ with $|L_z|\ll L$. On the other hand  the squeezed
configuration $|r_z|\ll|r_{\bot}|$ is to be formed with use of $n\sim
1/\Delta\theta$ different  polynomials with $(L-|L_z|)\ll L$.

 To consider these dynamical limits on light cone we first note that in
the (almost) 2+1 squeezed case there are two independent possibilities
to excite the system. The first one is to increase $|L_z|\approx L$
keeping the  radial quantum number to be constant.
As we will derive below this regime on light cone corresponds to the
transverse rotating string with the spectrum $M^2=2P_+H$
\be
M^2\to2\pi\sigma\cdot L,~~L\approx |L_z|\gg n_r
\ee
In the opposite case, $n_r \gg L\approx |L_z|$ the transverse linear
potential describes the excitations of $n_r$
\be
H\to \frac{1}{2P_+}(2|\vec{p}_{\perp}|+\sigma |\vec{r}_{\bot}|)^2
\ee
so that
\be
M^2\to 2\pi\sigma (2n_r) \; , \;\; n_r \gg L \approx |L_z|\gg 1
\ee

We emphasize that these regimes can be interpreted as the transverse
projections of the proper rest frame regimes. In contrast to that on the
light cone there exists a quasi $1+1$ dynamics of the stretched configuration
peculiar for this frame and described, as we will prove, by the well known
$1+1$ QCD Hamiltonian of t'Hooft  [17].
\be
H\to \frac{1}{2P_+}(2\sigma|P_+ r_-|), n_r\gg L\gg |L_z|
\ee
Longitudinal excitations of this configuration obviously correspond in the
rest frame to the increase of $n_r$. Taking into account the proper boundary
conditions one obtains
\be
M^2\to 2\pi\sigma(2n_r) \; , \;\; n_r \gg L \gg |L_z|
\ee

We can conclude that at least asymptotically one recovers the rest frame
form (3.11) of the spectrum, so that the regime (3.15) is the analog of the
rest frame string regime while the regimes (3.16) and (3.18) are the
counterparts of the rest frame potential dynamics.

Let us derive now the light cone asymptotics, discussed above.

We start with  stretched configuration (3.13) and consider first
 $1+1$ analog of our Hamiltonian (2.28).  To go over to the two
dimensional case one should omit everywhere transverse degrees  of
freedom; some care is needed for the $r^2_{\bot}$ terms in  the
denominators: one should keep in mind that $\nu$ is an auxiliary function
which can be redefined  since it is to be found from the condition of the
extremum (2.29).  Correspondingly we introduce $\bar{\nu}$ via
 \be
\bar{\nu}\equiv \nu /r^2_{\bot}
\ee
which stays nonzero, while $\nu$ and
$r^2_{\bot}$ tend to zero; putting  in eq. (2.28) $p_{\bot}=0$ and
$r_{\bot}=0$ one obtains the one-dimensional Hamiltonian \be
H_{long}=\frac{1}{2} \{ \frac{m^2_1}{\mu_1}+\frac{m^2_2}{\mu_2}+ \int
 \frac{\sigma^2}{\bar{\nu}} d\beta + \frac{\bar{\nu}_0 a_1
r^2_-}{\mu_1+\mu_2} \} \ee

 Here we used notation
 $$\bar{\nu }_0= \int^1_0 \bar{\nu}d\beta$$
 It is clear from (3.21) that $\bar{\nu}^{ext}$ does  not depend on $\beta$
 (since $\frac{\delta H}{\delta \nu(\tau,\beta)}$ does not depend on it),
 and we can rewrite \be H_{long}=\frac{1}{2} \{
  \frac{m^2_1}{\mu_1}+\frac{m^2_2}{\mu_2}+ \frac{\sigma^2}{\bar{\nu}} +
  \frac{\bar{\nu}P_+ r^2_-}{\mu_1+\mu_2} \} \ee

  Taking into account that $\nu\to 0$ at $r^2_{\bot}\to0$ one obtains
   \be P_+=a_1= \mu_1+\mu_2+\nu_0\to \mu_1+\mu_2 \ee
    Finally the
  Hamiltonian after substitution of the  extremum  value of $\bar{\nu}$
  takes the form \be H_{long}=
  \frac{m^2_1}{2\mu_1}+\frac{m^2_2}{2\mu_2}+ \sigma|r_-| \ee

  Expressing $\mu_1, \mu_2$ in terms of  the familiar
  $x$ parameter (2.19), $0\leq x \leq 1$
   \be
  P_+=p_{1+}+p_{2+}=\mu_1+\mu_2;~~ \mu_1=P_+x,~~\mu_2=P_+(1-x) \ee
  we get
   \be
  H= \frac{1}{2P_+}( \frac{m^2_1}{x}+\frac{m^2_2}{1-x})+ \sigma|r_-| \ee
  and   comparing with $H=\frac{M^2}{2P_+}$ one has
   finally \be M^2= \frac{m^2_1}{x}+\frac{m^2_2}{1-x}+ 2 \sigma| P_+ r_-|
   \ee
   Since $x$ is canonically
  conjugated to $P_+r_-$  (see (3.5) and
  (3.8)) equation (3.27) is to be considered as an operator equation for
  the wave function $\Psi(x)$ of a meson in the light-cone system \be (
  \frac{m^2_1}{x}+\frac{m^2_2}{1-x})\Psi(x)+ \int K(x,y)\Psi(y) dy=
 M^2\Psi(x)
  \ee
 The operator $K$ is easily obtained from the Fourier transform of  the
 last term in eq.(3.8) \be K(x,y)\Psi(y) =
 -\frac{\sigma}{2\pi}\int\frac{dy\Psi(y)}{(x-y)^2} \ee
  Our equation (3.28)
 with the integral kernel (3.29) coincides with the t'Hooft integral
 equation, obtained in the framework of the $1+1$ QCD for $N_C\to \infty$
 [17], when a proper identification of parameters is made
  (for details of this comparison see
 Appendix).

 A. Let us consider the 3+1 limit (3.19) when $n_r \gg L\gg |L_z|$ and
  the stretching condition (3.13) is satisfied. For this
 asymptotics one can keep only the following terms in the Hamiltonian
 (2.28)
 \be
 H=\frac{1}{2P_+}
 (\frac{(P_+ r_-)\int yd\beta}{r^2_{\bot}(1-\int yd\beta)}+ \sigma^2
 \int\frac{d\beta}{y}r^2_{\bot})
 \ee
Substituting the extremal values of $y(\tau,\beta)$
\be
y(\tau,\beta)=
 \frac{\sigma r_{\bot}^2}{|P_+r_-|+\sigma r^2_{\bot}}\to \frac{\sigma
 r^2_{\bot}}{|P_+ r_-|} \ll 1
 \ee
one obtains for the resulting Hamiltonian in the leading order
\be
H=\frac{1}{2P_+}(2\sigma|P_+ r_-|)
\ee

This is again the well known t'Hooft Hamiltonian for 1+1 QCD. In the 3+1 QCD
case it corresponds to the asymptotics (3.13) which leads to the following
spectrum of eq. (2.28) (see Appendix for details).
\be
M^2 =2\pi\sigma(2n_r) \; , \;\; n_r \gg L\gg |L_z|
\ee
We have identified here the longitudinal excitations of the system with the
radial quantum number excitations in agreement with consideration given
above.

We note here that in this regime the string contribution is dominant, but
corresponds to the light cone longitudinal linear potential. As a result
$y=\nu/P_+$ doesn't depend on parameter $\beta$ along the string and we
arrive at the dynamics which is the counterpart of the potential regime
(case 2) of the rest frame.

 B. Next we consider the case of configuration (3.14), when
   $L\approx|L_z|$, and  the radial quantum number of the Hamiltonian
   (2.28) $n_r$ is very large,
   \be n_r\gg L\approx|L_z| \ee
   so that condition
    (3.14) is satisfied. In this case one can keep in (2.28) only the fourth
       and the fifth terms on the r.h.s. and write
       \normalsize
       \be
       H=
\frac{1}{2P_+}\{ \frac{p^2_{\bot}}{x-\frac{1}{2}+\int\beta zd\beta}+
\frac{p^2_{\bot}}{\int(1-\beta)zd\beta-(x-\frac{1}{2})}+\sigma^2
r^2_{\bot}\int\frac{d\beta}
{1-z}\}\ee
\large
where $z(\tau,\beta) =1-y(\tau,\beta)=1-\frac{\nu(\tau,\beta)}{P_+};$

Extremum in $x$ and $z$ is achieved for  $x=\frac{1}{2}$ and
\be
 z(\tau,\beta)=
\frac{2|p_{\bot}|}{2|p_{\bot}|+\sigma|r_{\bot}|} \ee The leading term  in
(3.35) looks like
 \be H\approx
\frac{1}{2P_+}(2|p_{\bot}|+\sigma|r_{\bot}|)^2 \ee and yields the mass
spectrum
\be M^2=2\pi\sigma(2n_r),~~n_r=1,2,...  \ee
Here we have identified  the transverse radial excitations with the
excitations of radial quantum number $n_r$.

 In this potential like regime the string and the quarks make comparable
contribution to $M^2$. The string constitutes the transverse linear
potential, so that one has string energy density $\nu/P_+$ independent on
$\beta$ which corresponds to the case 2 of the rest frame.

C. Consider now the case of large $L\approx |L_z|$, which corresponds to the
configuration (3.15), discussed at the beginning of this Section. As in the
previous case, the motion is  almost in the transverse plane, both
 $p_{\bot}$ and $r_{\bot}$ are large and roughly
perpendicular to each other so that condition (3.14) is satisfied. The
dominant terms in (2.28) are
\be H\cong \frac{1}{2P_+}\{
\frac{L^2}{r^2_{\bot}\tilde{a}_3}+
\sigma^2\int\frac{d\beta}
{y}r^2_{\bot}\}\ee
where $\tilde{a}_3(y)=a_3/P_+$. We can find the extremal value of
$r^2_{\bot}$ from (3.39) and obtain the effective Hamiltonian of the
          transverse rotating string without quarks.  \be H_{eff}
=\frac{1}{2P_+}2|L|\sigma (\frac{1}{\tilde{a}_3}\int\frac{d\beta}
{y})^{1/2}\ee
          with the conditions that $a_3$ is given by (2.15) and $P_+=a_1$.

          The minimization of $H_{eff}$ with respect to $y(\tau,\beta)$
          under the constraint $a_1=P_+$ yields as
          in I for the limiting  case (3.15).
           \be y_{ext}
=\frac{\nu_0(\beta)}{M_0}=\frac{1}{M_0}(\frac{8\sigma L}{\pi})^{1/2}
\frac{1}{\sqrt{1-4(\beta-\frac{1}{2})^2}}
\ee
and
\be
H_{eff} = \frac{M^2_0}{2P_+},~~M_0^2= 2\pi\sigma L
\ee
Thus we recover for the configuration 2, where $|L_z|\approx L\gg n_r$, again
the correspondence with the rest frame spectrum (3.11).

In this string regime (case 1 of the rest frame) the main contribution comes
from the rotating transverse string and as a consequence $\nu/P_+$ does
depend on $\beta$ in accordance with Lorentz gamma factor (see  I  for
details).

 \setcounter{equation}{0}
\renewcommand{\theequation}{4.\arabic{equation}}

\section{Conclusions}

In this paper we have derived from the QCD Lagrangian the
 light--cone Hamiltonian, describing quarks connected by the string
  and analyse different dynamical regimes of the system.  The main
 assumption (supported by the cluster expansion arguments and lattice data
 [9]) is the area law for the Wilson loop.

 The resulting Hamiltonian (2.28) is both reasonable and unexpected.
 Reasonable, since the spectrum (estimated here asymptotically)
 coincides with the c.m. spectrum found in I, and agrees well with
 experimental meson spectra [18].

     The form of Hamiltonian is at the same time unexpected from point
  of view of the standard  light--cone theory developed heretofore. The main
  new point in our formalism is the considerable contribution of the
  string into the total momentum $P_+$ and orbital momentum $L^2$. Therefore
  the usual separation of the pure kinetic and pure potential part in the
  light--cone Hamiltonian
   \be
  H=\frac{2\sigma}{2P_+}(M^2_0+W) \ee
  where $M_0^2$ is the free particle  kinetic part, is not valid here. In
  our 3+1 case the string is not inert and interacts in the complicated way
  with quarks. In particular for the case (A) of $n_r\gg L\gg | L_z|$
  one obtains
  \be
   H\approx \frac{2\sigma}{2P_+}|P_+r_-|
   \ee
    so that $P_+$ enters $M^2$
  in the drastic way, which is unusual for the standard formulation [11,12].

  This fact is also supported by an independent argument -- a
comparison with the 1+1 QCD shows that the dominant term in the
t'Hooft's equation coincides with (4.2), as we have demonstrated  it
in Section 3.2 and Appendix.

The presence of nonperturbative string term like (4.2) is unexpected
also from the point of view of the light--cone quantization approach
[7], where nonperturbative contributions are missing. In this context
our Hamiltonian can be considered as a useful lesson how to
incorporate nonperturbative effects into the formalism.

To complete the light--cone formalism for spinless quarks connected by
the string one needs to construct  4 dimensional
tensor $M_{\mu\nu}$, total angular momentum $L^2$, check Poincare group
commutations between operators and compare c.m.
and light--cone wave functions. It is also clear, that the wave function
$\psi(x,p_{\bot})$ is to be found by numerical computations with the
Hamiltonian (2.28). This will allow numerous applications for formfactors,
structure functions etc.

The work in this direction is now in progress.

The authors acknowledge financial support of the Russian Fund for
Fundamental Research in the framework of the grant 93-02-14397.

We are grateful for useful discussions to A.S.Gors\-ki,
L.A.Kon\-dra\-tyuk,
I.M.Na\-ro\-dets\-ky and K.A.Ter-Martirosyan.
\newpage

\setcounter{equation}{0}
\renewcommand{\theequation}{A.\arabic{equation}}

 \begin{center}
 \large{Appendix }\\

 {\bf  Comparison of eq.(3.28) with the t'Hooft's equation in the $1+1 $
 QCD}

 \end{center}

 One can rewrite the potential term in (3.28) as
 \be
 V(r_-)\Psi(r_-) = \int \tilde{V}(p-q)\tilde{\Psi}(q) \frac{dq}{2\pi}
 \ee
 where the Fourier transform $\tilde{V}(p)$ is defined via
 \be
 \tilde{V}(p)= \int^{\infty}_{-\infty}{V}((r_-)e^{ipr_-}
 dr_-~,~~V(r_-)=2P_+\sigma|r_-|
  \ee
 For the integral in (A.2) to have sense, one has to introduce a cut-off
 $e^{-\gamma|r_-|}$ and finally put $\gamma\to 0$. With that from (A.2) one
 obtains
  \be
  \tilde{V}(p) = {2P_+\sigma}
   [\frac{1}{(\gamma-ip)^2}+\frac{1}{(\gamma+ip)^2}]
   \ee
   With the definition of the principal value integral as in [17] one has
   \be
   \tilde{V}(p) = - 4P_+\sigma P(\frac{1}{p^2})
   \ee
   where
   \be
   P\int \frac{\varphi(k) dk}{k^2}=\frac{1}{2}\int
   \frac{\varphi(k+i\gamma)dk}{(k+i\gamma)^2}+\frac{1}{2}
 \int\frac{\varphi(k-i\gamma)dk}{(k-i\gamma)^2}
 \ee
 Insertion of (A.4)-(A.5) into (A.1) with the definitions
\be
 p=P_+x~;~~ q=P_+y
 \ee
 allows finally to represent (A.1) as
 \be
 V(r_-)\Psi(r_-)= -\frac{2\sigma}{\pi}P\int\frac{\tilde{\Psi}(y)dy}{(x-y)^2}
 \ee
 Eq.(A.7) coincides with the t'Hooft's kernel [17].

 A special attention should be given to the mass terms in the  full
 t'Hooft's equation
 \be
  M^2\varphi(x)=(\frac{\bar{m}^2_1}{x}+\frac{\bar{m}^2_2}{1-x})\varphi(x) -
   P(g^2/\pi)\int^1_0 \frac{\varphi(y)dy}{(y-x)^2}
    \ee
     where
     \be
    \bar{m}^2_i=m^2_i-\frac{g^2}{\pi}
    \ee
    and $m_i$ are the current quark masses.

    In the $1+1$ QCD at  $N_c\to \infty$  the mass shift $(-g^2/\pi)$ is due
     to the infrared region contribution of the
 selfenergy loop graphs, and the negative sign is
 due to attractive Coulomb (confinement)
 interaction $g^2|r_-|$. The situation is
 different in the $3+1 $ QCD, where the same loop
 graph is only ultraviolet divergent and yields a usual renormalization of
 the quark propagator.  Therefore there the current quark mass is defined at
 some normalization scale (usually at $\mu = 1 GeV$) and is ascribed the
 known  value (e.g. $m_u=4 MeV, m_d= 7 MeV$ etc). However due to chiral
 symmetry breaking there appears a chiral mass of quark in the form of a
 nonlocal gauge-covariant operator $\hat{M}(x,y)$ [19].

 At large distances, $R\gg\rho$ ( where $\rho$ is
 the average instanton size, $\rho \approx 0.2
 \div 0.3 fm$) one can approximate $\hat{M}$ by a
 local gauge-invariant mass $M_{ch}$, which
 finally enters into the Hamiltonian (I. 73) via
 $\sqrt{\vec{p}^2+(M_{ch}+m)^2}$.
 The spectrum of (A.8) is calculated in [17] and at large $n$ the WKB
 approximation yields [17]
 \be
 M^2_n = g^2\pi n + (\bar{m}^2_1+\bar{m}^2_2) ln n +C
\ee

The constant is close to $g^2\frac{3}{4} \pi$.

The lowest mass values for $m_1=m_2=0$, are approximately
$$M_0^2=0;~~ M^2_1 \cong 0.6 \frac{g^2}{\pi};~~ M_2^2\cong 1.43
\frac{g^2}{\pi}$$

To compare with our eq.(3.28), one should replace $g^2\to 2\sigma$. The
spectrum (A.10) is to be compared with that of relativistic quark model [2],
which asymptotically is
 \be M^2_n= 4\pi \sigma n + 2m^2 + 4m^2 ln
\frac{M_n}{m} +m^2_0 \ee
with $m^2_0 = 3\pi\sigma$. Taking into account that for 3+1 case only odd
solutions $\varphi(r_-)=-\varphi(r_-)$ (and even $n$ in eq. (a.10) )are
relevant one can see a close correspondence of (A.10) and (A.11) up to
a constant $2m^2$ .

The wave functions $\varphi(x)$, solutions of (A.8) behave at the boundary
$x=0$ as $\varphi(x) \sim x^{\beta_1}$, with $\pi\beta_1
ctg\pi\beta_1+\alpha_1=0$ [17]. Boundary conditions are
\be
\varphi_n(0)= \varphi_n(1)=0,~~ n=1,2,...
\ee
Approximately $\varphi_n(x) \approx sin  n\pi x,~~ n=1,2,...$
For $m_1=m_2=0$ there appears a ground state $n=0$ with $M_0=0$ with wave
function $\varphi(x) \equiv 1$, which has no counterpart in $3+1$.

\newpage
\begin{center}

{\bf Reference}

\end{center}

\begin{enumerate}
\item {\it   A.Yu.Dubin, A.B.Kaidalov and Yu.A.  Simonov}// Yad.
Fiz. 1993. V. 56 p.12, HEP-PH 9311344
 \item{\it  J.Carlson et al.}// Phys.  Rev.  1983. V. D27. P.233\\
 {\it J.L.Basdevant and S.Boukraa}// Z.Phys. 1986. V.C 30. P. 103
 \item{\it A.Yu.  Dubin, A.B.Kaidalov and Yu.A.Simonov}// Phys.  Lett.
 1994. V. B323. P.41
 \item{\it  D.La Course and M.G.Olsson}// Phys.Rev.  1989. V. D39. P. 2751
 \item{\it A.H.Mueller}// Phys.Rep . 1981. V. 73. P. 237\\
  {\it L.V.Gribov, E.M.Levin and M.G.Ryskin}// Phys.Rep. 1983 V. 100. P. 1.
   \item{\it R.P.Feynman}// Photon-Hadron Interactions,
  W.A.Benjamin, Reading, 1972.
   \item{\it S.J.  Brodsky and H.-C.Pauli}//
  Lectures at the 30th Schladming Winter School in Particle Physics,
  SLAC-PUB-5558 (1991)
  \item{\it  Yu.A.Simonov}// Nucl.  Phys. 1988. V. B307.P. 512
 \item{\it  Yu.A.Simonov}// Yad. Fiz. 1991. V. 54. P. 192
 \item{\it  Yu.A.Simonov}// Phys. Lett.1989. V. B226. P. 151
 \item{\it P.A.M.Dirac}// Rev.Mod. Phys. 1949. V. 21. P. 392
 \item{\it  M.V.Terentiev}// Yad.  Fiz.  1976. V.  24. P. 207\\
 {\it V.B.Berestetsky and M.V.Terentiev} //Yad.Fiz.  1976. V. 24. P.1044
 \item{\it  Yu.A.Simonov}// Perturbative Theory in the Nonperturbative
 QCD vacuum, HD-THEP-93-16; HEP-PH-9311247
 \item{\it  Yu.A.Simonov}// Nucl.
 Phys. 1989. V. B324. P. 67
  \item{\it  V.Marquard and H.G.Dosh}//
 Phys.  Rev. 1987. V. D35. P. 2238
 \item{\it  E.L. Gubankova and
 A.Yu.Dubin}// Phys. Lett. B in press.
  \item{\it  G.t'Hooft}// Nucl.Phys. 1974. V.B75. P. 461.
 \item{\it Yu.A.Simonov}// Talk at the Hadron-93 conference in Como, June
 1993, HEP-PH 9311217
 \item{\it  Yu.A.Simonov}// Yad Fiz. 1994. V. 58. P.N 8, HEP-PH 9401320
 \item{\it V.L.Chernyak and A.R.Zhitnitsky}// Phys.Rep. 1984. V.112. P.174

    \end{enumerate}

   \end{document}